\documentclass[english,aip,pof,groupedaddress,preprint,floatfix,showpacs,showkeys,longbibliography]{revtex4-1}

\usepackage{graphicx}
\usepackage{dcolumn}
\usepackage{amsmath}
\usepackage{epstopdf}
\usepackage{graphicx}
\usepackage{hyperref}
\usepackage{color}
\usepackage[english]{babel}
\usepackage{lmodern}
\usepackage[T1]{fontenc}

\usepackage[normalem]{ulem}

\makeatother

\begin{document}

\title{Effects of Orthogonal Rotating Electric Fields on Electrospinning Process}

\author{M.~Lauricella$^{1}$}
\email[First author: ]{m.lauricella@iac.cnr.it}
\author{F.~Cipolletta$^{1}$}
\author{G.~Pontrelli$^{1}$}
\author{D.~Pisignano$^{2,3}$}
\author{S.~Succi$^{1,4}$}
\email[Corresponding author: ]{s.succi@iac.cnr.it}
\affiliation{$^1$Istituto per le Applicazioni del Calcolo CNR, Via dei Taurini 19, IT 00185 Rome, Italy}
\affiliation{$^2$Dipartimento di Matematica e Fisica ``Ennio De Giorgi'', University of Salento, via Arnesano, IT 73100 Lecce, Italy}
\affiliation{$^3$NEST, Istituto Nanoscienze-CNR, Piazza S. Silvestro 12, IT 56127 Pisa, Italy}
\affiliation{$^4$Institute for Applied Computational Science, Harvard ``John A. Paulson'' School of Engineering and Applied Sciences, Cambridge, MA 02138, United States}


\begin{abstract}
Electrospinning is a nanotechnology process whereby an external electric field is used to accelerate 
and stretch a charged polymer jet, so as to produce fibers with nanoscale diameters. 
In quest of a further reduction in the cross section of electrified jets hence of a 
better control on the morphology of the resulting electrospun fibers, we explore
the effects of an external rotating electric field orthogonal to the jet direction. 
Through intensive particle simulations, it is shown that 
by a proper tuning of the electric field amplitude and frequency, a reduction of 
up to a $30 \%$ in the aforementioned radius can be obtained, 
thereby opening new perspectives in the design of future ultra-thin electrospun fibers. 
Applications can be envisaged in the fields of nanophotonic components as well as 
for designing new and improved filtration materials.
\end{abstract}


\maketitle

\section{Introduction}\label{sec:01}

Electrospinning has witnessed a dramatic upsurge of interest in recent
years because of its potential to produce ultra-fine fibers with 
sub-micrometer diameters 
(see Refs ~\cite{li2004electrospinning,greiner2007electrospinning,
carroll2008axisymmetric,huang2003review, yarin2014fundamentals, 
wendorff2012electrospinning,pisignano2013polymer}). 
Though routinely realizable in the laboratory, electrospinning is a 
complex phenomenon to analyze because of the
coupling between the electric field and 
the non-linear deformation of the fluid, the latter 
being dictated by the rheology of the material.
As a consequence, the resulting jet (fiber) diameter is
affected by several material, design, and operating parameters.

In this context, computational models represent 
a useful tool to investigate the underlying physics
of electrospinning and provide information which may be used 
for the design of new electrospinning experiments and nanofibrous materials. 
Several strategies have been pursued to model the process, 
which can be broadly classified within two main families, Eulerian and Lagrangian. 
The former is based on a fixed-grid discretization of the partial differential
equations of continuum fluid-dynamics \cite{yarin2001bending,hohman2001electrospinning,ganan1997theory,huebner1971instability}, while in the latter the grid moves
with the flow, taking the form of particle-like ordinary differential 
equations.\cite{reneker2000bending,lauricella2016three,carroll2011discretized,ganan1994electrostatic}

\begin{figure}[!hbtp]
\includegraphics[width=0.9\hsize,clip]{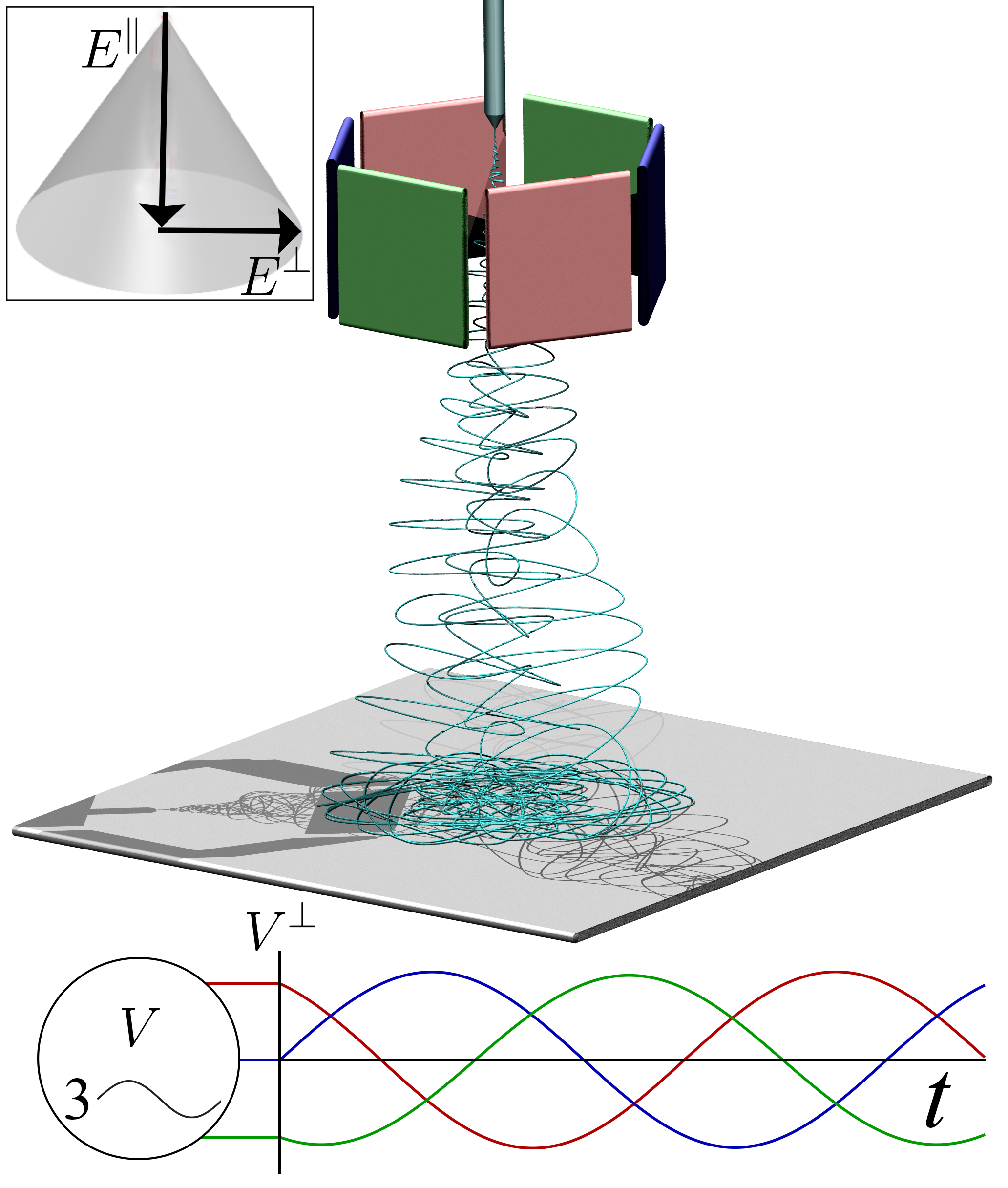}
\caption{Sketch representation of the electrospinning process in presence of an orthogonal rotating
electric field (OREF). The jet, ejected from a nozzle, is stretched by an electric field $E_{\parallel}$ parallel to the $x$ axis. 
The OREF $E_{\perp}$ can be generated
by a series of capacitor plates hexagonally arranged, and connected to a
three-phase power source.
The electrical power source is graphically represented in the bottom part of the figure. Here, the three
phase voltage show colors corresponding to the color of the connected capacitor plate.}
\label{fig01}
\end{figure}

By using suitable theoretical models, the effects of the 
parameters on the fiber diameter can be
systematically studied and assessed, both analytically and numerically. 
For example, it has been shown that bending (or whipping) instabilities of 
electrical and hydrodynamical nature, are mostly
responsible for jet stretching during the electrospinning 
process.~\cite{yang2014crossover,reznik2010capillary,reneker2000bending}
This behavior leads to a reduction of the cross section 
radius of electrospun nanofibers. 
In other studies, the attention is mostly focused on the 
morphological aspects, revealing a wide variety of 
pattern depositions of electrified jets.~\cite{yang2014crossover,reneker2008electrospinning} 
In the literature, one can find theoretical models to describe the jet 
dynamics and control of the fiber diameter, 
through numerical simulations based
on multi-parameter choice, involving the perturbation at the nozzle, the intensity of the fixed 
electric field, the density of the polymer  
solution.~\cite{reneker2008electrospinning,fridrikh2003controlling,reneker2000bending}
However, investigating new strategies to improve the overall control on electrospun nanofibrous 
materials, still is an open scientific and technological challenge. 
In the present work, we consider the effect of a rotating electric 
field orthogonal to the main electric field (Figure~\ref{fig01}).
In particular, we present a theoretical model and ensuing numerical simulations with
the JETSPIN code~\cite{lauricella2015jetspin,lauricella2016dynamic}, 
in order to identify the optimal values of the amplitude and rotational 
frequency of the orthogonal rotating electric field (OREF), which permit to
alter fiber morphology.

\begin{table}[!hbtp]
\caption{Parameters values in the simulations.}
\begin{ruledtabular}
\begin{tabular}{lll}
simulated time & $0.1$ & $s$ \\
discretization length step $l_{step}$ & $0.02$ & $cm$ \\
initial jet radius $R_0$ & $5 \cdot 10^{-3}$ & $cm$ \\
charge density & $4.4 \cdot 10^4 $ & $g^{\frac{1}{2}} \, cm^{-\frac{3}{2}} \, s^{-1}$ \\
fluid viscosity $\mu$  & $20$ & $g \, cm^{-1} \, s^{-1}$ \\
elastic modulus $G$  & $5 \cdot 10^4$ & $g \, cm^{-1} \, s^{-2}$ \\
collector distance $h$ & $16$ & $cm$ \\
external electric potential ($h E^{\parallel}$) & $30.021$ & $g^{\frac{1}{2}} \, cm^{\frac{1}{2}} \, s^{-1}$ \\
surface tension $\alpha$ & $21.13$ & $g \, s^{-2}$ \\
bulk velocity $\upsilon_s$ & $0.28$ & $cm \, s^{-1}$ \\
perturbation frequency $\omega_{pert}$ & $10^4$ & $s^{-1}$ \\  
perturbation amplitude $A_{pert}$ & $10^{-3}$ & $cm$ \\
OREF modulus $A$ & $\left[ 0.0, 10.0 \right]$ & $g^{\frac{1}{2}} cm^{-\frac{1}{2}} s^{-1}$ \\
OREF frequency $\omega$& $\left[ 0.5, 20.0 \right]$ & $10^4 \, s^{-1}$ 
\end{tabular}
\end{ruledtabular}
\label{tab:INPUT}
\end{table}

\section{The model}\label{sec:02}

The jet emitted by the nozzle is modeled by a 
finite set of $n$ parcels at a distance $l_i$
(distance between the $i$th and $(i+1)$th parcel), connected via viscoelastic 
elements, similarly to previous electrospinning models.\cite{reneker2000bending,kowalewski2005experiments,sun2010three,carroll2011discretized}
Each jet parcel represents a cylindrical element of jet volume $V_i=V_{0}$ and initial height $l_{i}=l_{step}$ 
(initial length step of jet discretization). 
As a consequence, the initial radius $R_{0}$ of the jet element 
is equal to $\sqrt{V_{0}/\pi l_{step}}$. 
From this representation, the following set of 
equations of motion (EOM) can be written 
for each parcel $i$ (hereafter, we shall consider the $x$-axis 
pointing from the nozzle to the collector):

\begin{align}
&\frac{d \vec{r_i}}{d t} = \vec{v_i}, \label{eq01} \\
& \nonumber \\
&\frac{d \sigma_i}{d t} = \frac{G}{l_i} \frac{d l_i}{d t} - \frac{G}{\mu} \sigma_i, \label{eq02} \\
& \nonumber \\
&m_i \frac{d \vec{v}_i}{d t} = q_i \vec{E} + \sum_{ j \neq i} \left( \frac{q_i q_j}{\vert \vec{r}_j - \vec{r}_i {\vert}^{2}} \vec{u}_{i,j} \right) - \pi {R_i}^2 \sigma_i \vec{t}_i  \nonumber \\
& + \pi {R_{i+1}}^2 \sigma_{i+1}  \vec{t}_{i+1}
+ k_i \, \pi \left( \frac{R_i + R_{i+1}}{2} \right)^2 \alpha \, \vec{c}_i, \label{eq03}
\end{align}

In the above, the subscript $i$ stands for the $i$-th parcel, $\vec{r}_i$ is the position vector, $\vec{v}_i$ is the velocity vector, 
$G$ is the elastic modulus, $\mu$ is the viscosity of the fluid jet, $\sigma$ is the stress, 
$R$ is the cross sectional radius, $q$ is the charge, $\vec{E}$ is the electric field (Figure~\ref{fig01}), 
$\vec{u}_{i,j}$ is the unit vector from parcel $i$ to parcel $j$, $\vec{t}_i$ is the unit vector pointing from 
parcel $i$ to $(i-1)$, $k_i$ is the local curvature, $\alpha$ is the surface tension coefficient 
and $\vec{c}_i$ is the unit vector pointing from the parcel $i$ to the local centre of curvature.
It is worth stressing that the filament radius $R_{i}$ at each $i$-th parcel location is equal to  $\sqrt{V_{0}/\pi l_{i}}$,
as a result of the volume conservation.
Note that the constitutive Eq \ref{eq02} models a Maxwell material with constant viscosity, 
in line with the approach of Refs \cite{reneker2000bending,kowalewski2005experiments}.

The aforementioned system of EOM is numerically solved, 
starting each simulation with only two bodies:
a parcel fixed at $x=0$, representing the
spinneret nozzle, and a second parcel modeling the initial jet segment 
located at distance $l_{step}$ from
the nozzle along the $x$ axis with a given initial velocity $ \upsilon^{\circ}_{i}$ (defined below).  
Once the last parcel reaches a distance  $2 l_{step} $ away from the
nozzle, a new parcel (third body) is placed at a distance $l_{step}$, 
the length step used to discretize the jet as a sequence of parcels.
Repeating this injection rule over time, we
obtain a set of $n$ discrete jet elements.

The initial velocity is defined as 
$\upsilon^{\circ}_{i}=\upsilon_{s}+\upsilon_{d,i}$,
where $\upsilon_{s}$ is a velocity term along the $x$ axis, modelling 
the bulk fluid velocity in the syringe needle, and $\upsilon_{d,i}$ 
denotes the dragging velocity
\begin{equation}
\upsilon_{d,i}=\frac{\upsilon_{i-1}-\upsilon_{s}}{2}.
\end{equation}
The extra term $\upsilon_{d,i}$ accounts for the drag effect of the electrospun 
jet on the last inserted segment ($i-1$). Note that the definition of $\upsilon_{d,i}$
was chosen in such a way as to keep the velocity strain unchanged 
before and after the parcel insertion.

Furthermore, we take into account a periodic nozzle perturbation with 
frequency  $\omega_{\text{pert}}$ and amplitude $A_{\text{pert}}$, which models fast mechanical oscillations nearby the
spinneret. This perturbation results 
in the emission of a conic helix jet. 
We are interested in adding an OREF to the above configuration, namely:
\begin{equation}
\label{eq05}
\vec{E} = \vec{E}^{\parallel} + \vec{E}^{\perp}.
\end{equation}
Hereafter, the main electric field and the OREF are denoted it by $\vec{E}^{\parallel} = \left( E_x,0,0 \right)$ and 
$\vec{E}^{\perp} = \left( 0, E_y,E_z \right)$, respectively.
In equations: 
\begin{eqnarray}
\label{eq06}
E_y (A,{\omega} ,t) = A \cos \, \omega t , \nonumber \\    
E_z (A,{\omega} ,t) = A \sin \, \omega t ,   
\end{eqnarray} 
where $A (\text{g}^{\frac{1}{2}} \, \text{cm}^{-\frac{1}{2}} \text{s}^{-1})$ is the modulus, $\omega (s^{-1})$ is the angular frequency.
OREFs have already been treated in the literature, in the context of plasma confinement by means of a series of 
capacitor plates with alternating current around the apparatus.~\cite{huang1997steady}
In JETSPIN, we modified the EOM according to equation~(\ref{eq05})--(\ref{eq06}).

With the exception of $A$ and $\omega$, in all simulations we use the
numerical parameters proposed in Ref. \cite{lauricella2015jetspin}.
These values were assessed by comparing
with experimental data from an electrospinning process of 
polyvinylpyrrolidone (molecular weight = 1300 kDa, mixture of 
ethanol and water 17:3 v:v, with PVP concentration 2.5 wt\%).\cite{montinaro2015sub} 
In particular, we consider
the surface tension $\alpha=21.1\, \text{g}\;  \text{s}^{-2}$ from Ref. \cite{yuya2010morphology},
the elastic modulus $G=5 \cdot 10^{4} \, \text{g} \; \text{cm}^{-1} \text{s}^{-2} $ from Ref. \cite{morozov2012water},
and the shear
viscosity $\mu=100 \, \mu_{0}$, where $\mu$ 
is taken equal to two orders of magnitude larger than the zero-shear viscosity 
$\mu_{0}=0.2\, \text{g} \; \text{cm}^{-1} \text{s}^{-1}$, see Ref. 
\cite{yuya2010morphology,buhler2005polyvinylpyrrolidone}. This is because the strong longitudinal 
flows we are dealing with, can lead to an increase of the extensional viscosity, as observed in the literature.\cite{reneker2000bending,yarin1993free}
Finally, $\upsilon_{s}$ is taken equal to 0.28 $\text{cm} \; \text{s}^{-1}$ which corresponds to 
a constant flow rate of $ 2 \text{mL} \; \text{h}^{-1}$ in a needle of radius $250 \mu \text{m}$.
The values of the simulation parameters are summarized in Table \ref{tab:INPUT}.


\section{Results}

In the following, we present an exploratory study of the effects of the
amplitude $A$ and frequency $\omega$ on the electro-spinning process.
We wish to emphasize that this does not represent a fully-fledged analysis of
the non-linear dynamical behaviour of this complex system, but rather a computational
identification of the most interesting regions in the $A-\omega$ parameter space.

\subsection{Modulus and frequency of OREF}

We investigate the effect of $\vec{E}^{\perp}$ in  equations~(\ref{eq06}), 
compared to the standard case  $\vec{E}^{\parallel}$, and study the way that 
the jet morphology is deformed for different choices of the free-parameters. 
We take $A^*=\vert \vec{E}^{\parallel} \vert \simeq 1.8763 \, \text{g}^{\frac{1}{2}} \, \text{cm}^{-\frac{1}{2}} \text{s}^{-1}$ 
and $\omega^*=\omega_{\text{pert}} = {10}^{4} \, \text{s}^{-1}$ 
as a reference values, respectively for $A$ and $\omega$ in equations~(\ref{eq06}).  
A wide range of physically relevant values of these parameters is spanned by selecting 
$A \in \left[ 0,10 \right]$ and $10^{-4} \omega \in \left[ 0.5, 20 \right]$ (see Ref. \cite{montinaro2015sub}). 
Such high values of the angular frequency $\omega$ mediate the OREF along the circumference 
orthogonal to $\vec{E}^{\parallel}$. Note that the dominant component of the force acting on each parcel 
is the one given by $\vec{E}^{\parallel}$, since the jet 
travels towards the collector without undergoing 
any breakup, such as that reported in literature.~\cite{yang2014crossover}

\begin{figure}[!hbtp]
\includegraphics[width=0.9\hsize,clip]{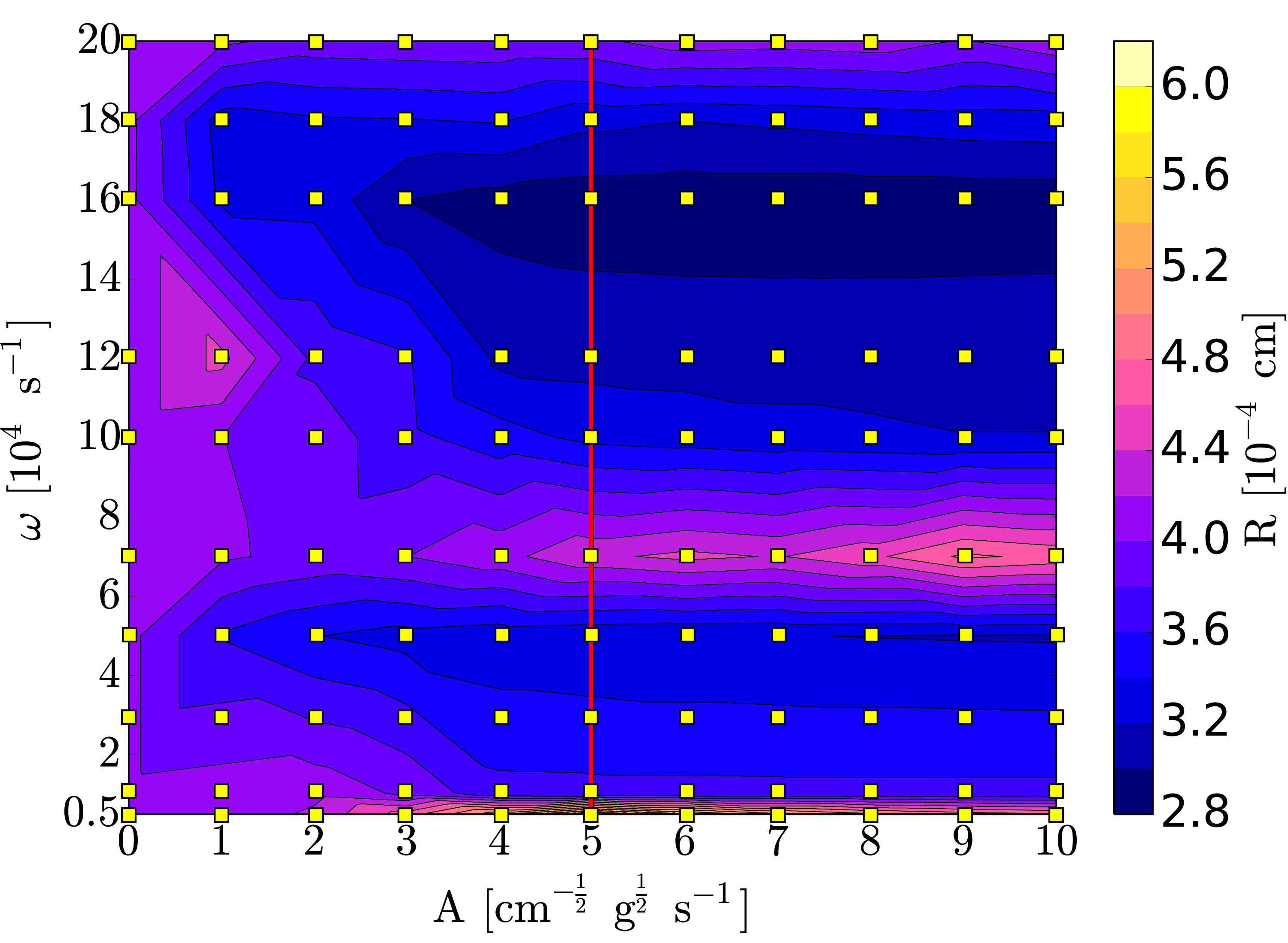}
\caption{Contour lines showing the simulated mean jet radius $R$ at the collector
as a function of $A \in \left[ 0,10 \right]$ and $\omega \in \left[ 0.5, 20 \right]$ parameters.
The $R(A,\omega)$ surface was assessed by a linear interpolation over a grid of pairs
of $A$ and $\omega$, shown in figure as yellow square symbols.}
\label{fig02}
\end{figure}

\begin{figure}[!htb]
\centering
\includegraphics[width=0.9\hsize,clip]{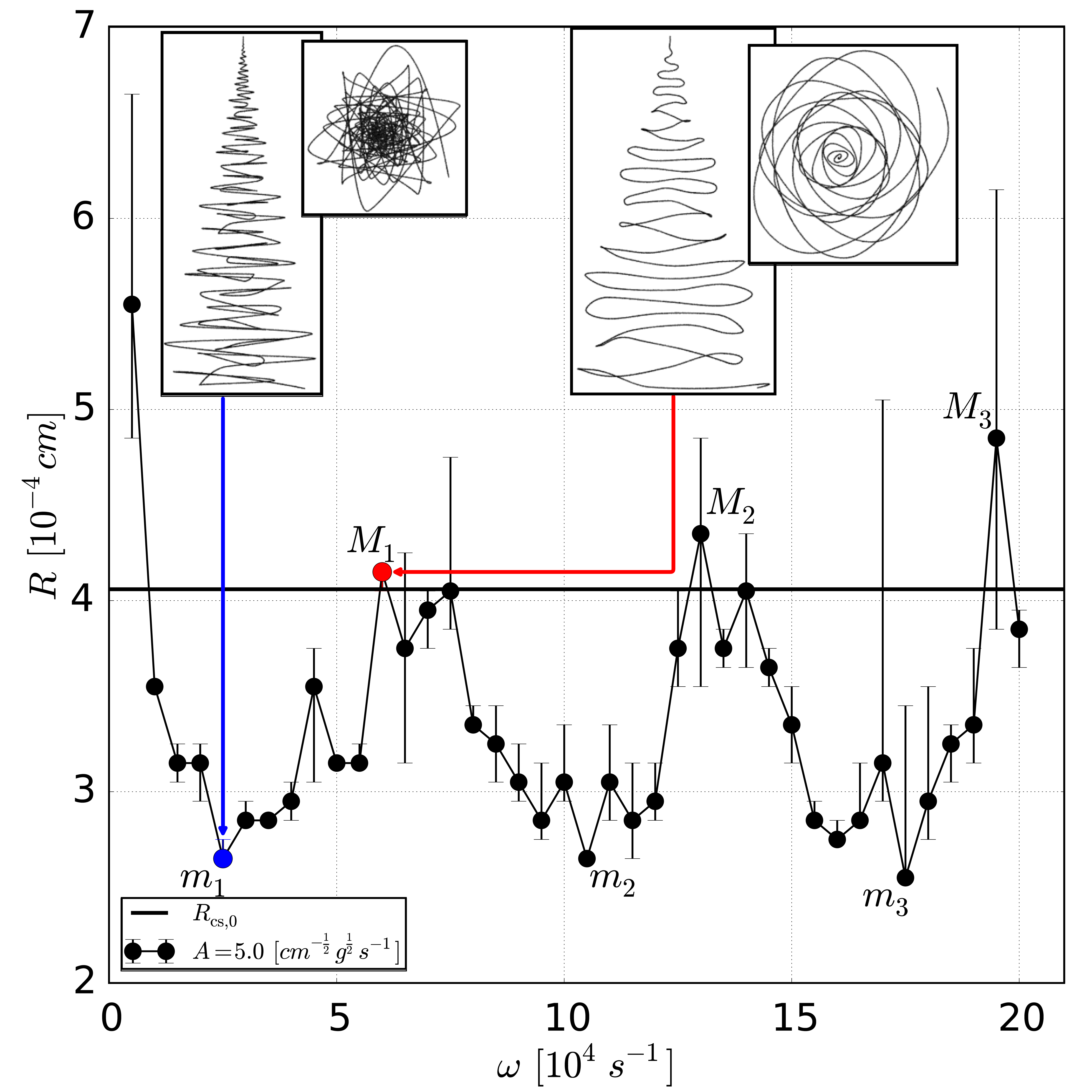}
\caption{Mean of cross section radius $R$ at the collector for $A = 5$ at 
several values of $\omega$, plotted with their respective interquartile as 
confidence intervals. The thick solid black horizontal line represents $R_{0}$, i.e. 
the mean value of $R$ in a standard electrospinning process (without OREF). 
The first minimum and maximum of $R$ are highlighted in red, with the insets 
showing the trajectories from the lateral and bottom planes.}
\label{fig03}
\end{figure}

The two components of the electric field are shown in the top left panel of 
Figure~\ref{fig01}, where the red and blue arrows
represent $\vec{E}^{\parallel}$ and $\vec{E}^{\perp}$, respectively, 
while the total field $\vec{E}$ runs over the transparent grey conic surface. 
We run several simulations, each for a total time of $0.1 s$.
After an initial drift $t_{\text{drift}} = 0.02$ s, the jet dynamics 
is assumed regular so that we can
estimate mean values of suitable observables over their stationary
statistical distributions. ~\cite{lauricella2015jetspin}
In particular, the mean value of the filament 
radius $R$ was estimated at the collector in different simulations with
several pairs of $A$ and $\omega$ values.
The surface of the mean
radius $R$ was reconstructed as a function of $A$ and $\omega$ by linear interpolation, 
shown by contour lines in Figure ~\ref{fig02}.
It is worth observing that for $A = 0$ the mean cross section 
radius reduces to $R_0 = 4.05 \cdot 10^{-4}$ cm,
in agreement with the
mean cross section values obtained for the standard 
case, $ \vec{E} = \vec{E}^{\parallel}$. 
Further, as a representative value, we analyze the case $A = 5 \, \text{g}^{\frac{1}{2}} \, \text{cm}^{-\frac{1}{2}} \text{s}^{-1}$ in Figure~\ref{fig03}, 
namely the red vertical line in Figure~\ref{fig02}. 
Along such a line,  we plot  $R$ as a function of $\omega$, 
together with the corresponding confidence interval 
(namely the interquartile interval). In this figure, 
the thick black horizontal line indicates
$R_{0}$ allowing a direct comparison between the cases with and 
without OREF. 

\begin{figure}[!htb]
\centering
\includegraphics[width=0.9\hsize,clip]{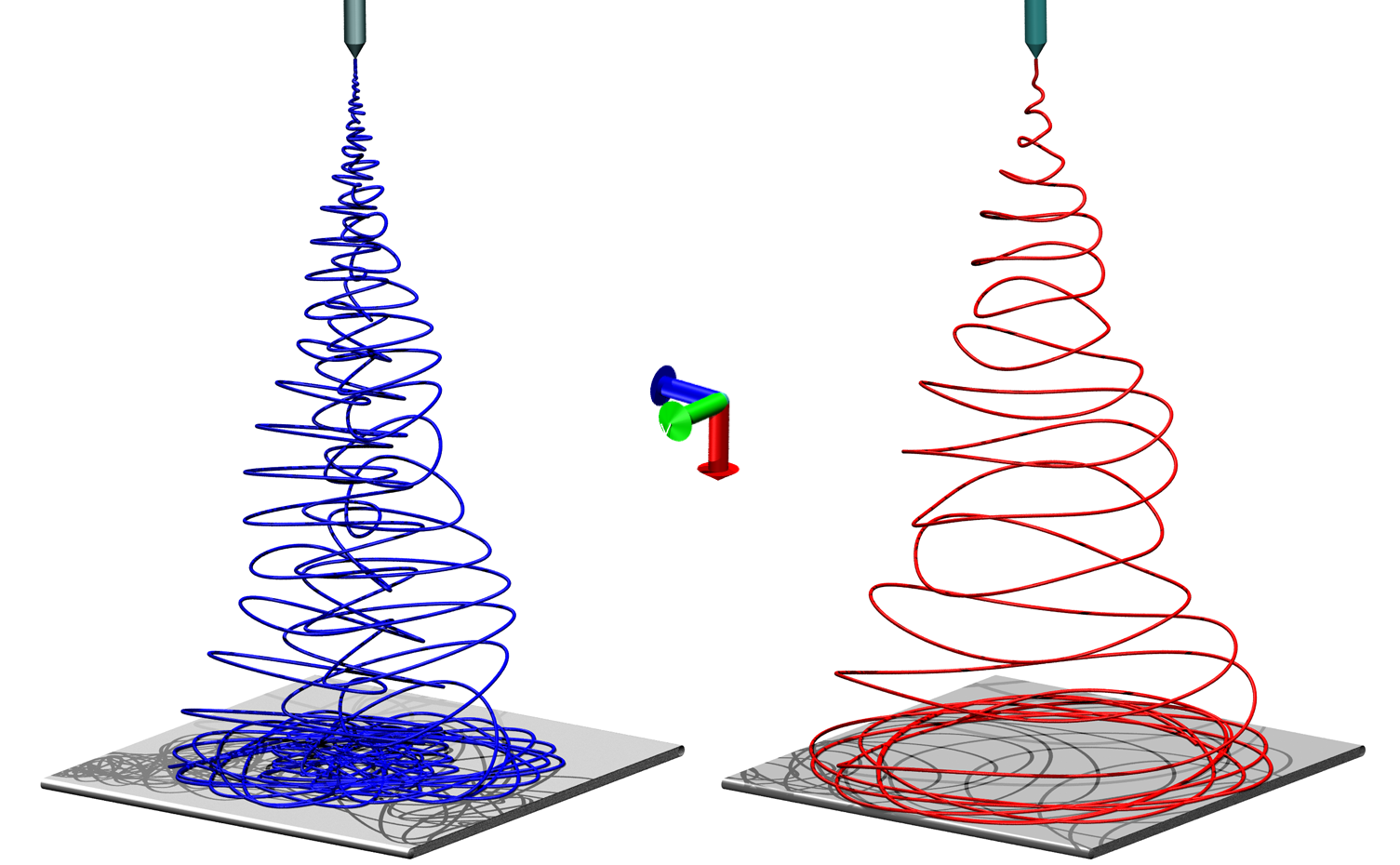}
\caption{Jet trajectories with $A = 5$, $\omega(m_{1}) = 2.5 \cdot {10}^4 \, \text{s}^-1$ 
(left blue curve) and $\omega(M_{1}) = 6 \cdot {10}^4 \, \text{s}^-1$ (right red curve). 
Axis convention, shown in the center, is as follows: red for $x$ axis, green 
for $y$ axis and blue for $z$ axis. While the red curve is  wider than the 
blue one, the latter appears more twisted, thus producing a thinner  $R$ 
(cfr with fig. 2 and 3).}
\label{fig04}
\end{figure}

From this plot, an oscillatory behavior of $R$ as a function of $\omega$ 
is clearly recognizable. 
Two representative points are singled out,  namely the first 
minimum $m_{1}$, with $R(m_{1}) = 2.65 \cdot {10}^{-4}$ cm, corresponding 
to  $\omega(m_{1}) = 2.5 \times 10^4 s^{-1}$ and the relative 
maximum $M_{1}$ with $R(M_{1}) = 4.15 \cdot {10}^{-4}$ cm, with $\omega(M_{1}) = 6 \times 10^4 s^{-1}$ . 
The corresponding jet-paths are reported in the
sub-panels: one is the view from the side and the other is the view 
from the collector, looking up to the
nozzle. It is worth noting that, although the helix associated  
with $\omega(M_{1})$ is wider than the one for
$\omega(m_{1})$, the latter is more entangled, meaning that the jet undergoes
longer-lived instabilities, resulting in a smaller value of $R$. 
3D representations of the two highlighted trajectories are displayed in 
Figure~\ref{fig04}, with color convention stated in the caption of Figure~\ref{fig03}. 
A reduction of the cross-section by about $34 \%$ with respect to the case 
without OREF (namely $A=0$ in Figure~\ref{fig02} for which $R_0$ is reported)
is observed.
It is worth observing that such a reduction is obtained without
altering the rheological properties of the jet by, say, changing 
the polymer concentration.       
This is comparable with similar results obtained by blowing-assisted 
electrospinning, where a gas stream, provided by suitable
distributors around the nozzle, is employed as additional stretching force.\cite{wang2005formation,lin2008preparation,hsiao2012effect}
The latter technique was recently extended by Sinha-Ray {\it et Al.} \cite{sinha2013supersonic} to include 
supersonic blowing gaseous stream in electrospinning
producing ultra-thin nanofibers.


Even though the functional relation $ R = R(\omega)$ is the result of a 
highly complex structural dynamics,
the oscillatory structure of such relation is relatively regular, 
and suggestive of a sort of resonant mechanism
underlying the OREF setup, which is likely to depend on
the frequency $\omega_{noz}$ of the mechanical oscillation of the nozzle. 
Here, we have kept $\omega_{noz}$  at a fixed value, because it
is practically easier to change the OREF frequency than modifying the 
inherent mechanical oscillations of the nozzle.

Nonetheless, we repeat the simulations corresponding to the two points
$m_{1}$ and $M_{1}$ in Figure \ref{fig03}, with a nozzle perturbation 
$\omega_{noz} =2.5 \cdot {10}^4 s^{-1}$. 
In both cases, we observe a change in the mean
cross section radius $<R>$ between $5$ and $10 \%$, compared to the previous
values, confirming that both frequencies $\omega$ and $\omega_{noz}$
contribute to the oscillatory jet dynamics.
At the moment, we have no clear theoretical explanation for such
oscillatory behavior, which depends on the overall non-linear dynamics
of the jet.
However, we can define suitable observables which help elucidating the 
effects of the OREF on the jet morphology (see next Subsection).

\subsection{Statistical analysis and overlap of trajectories}\label{sec:03b}

Here, we provide both statistical and time-dependent analysis performed over the trajectories related to the most
relevant $\omega$ values identified in the previous Subsection.
Each jet trajectory is the result of a complex dynamics, which presents 
an initial drift in the time lapse  
where the filament has not yet reached the collector. 
After such time lapse, the trajectory regularizes and consequently 
it becomes possible to 
analyze the  statistical distribution of the cross section radius.

\begin{figure}[!htb]
\centering
\includegraphics[width=0.8\hsize,clip]{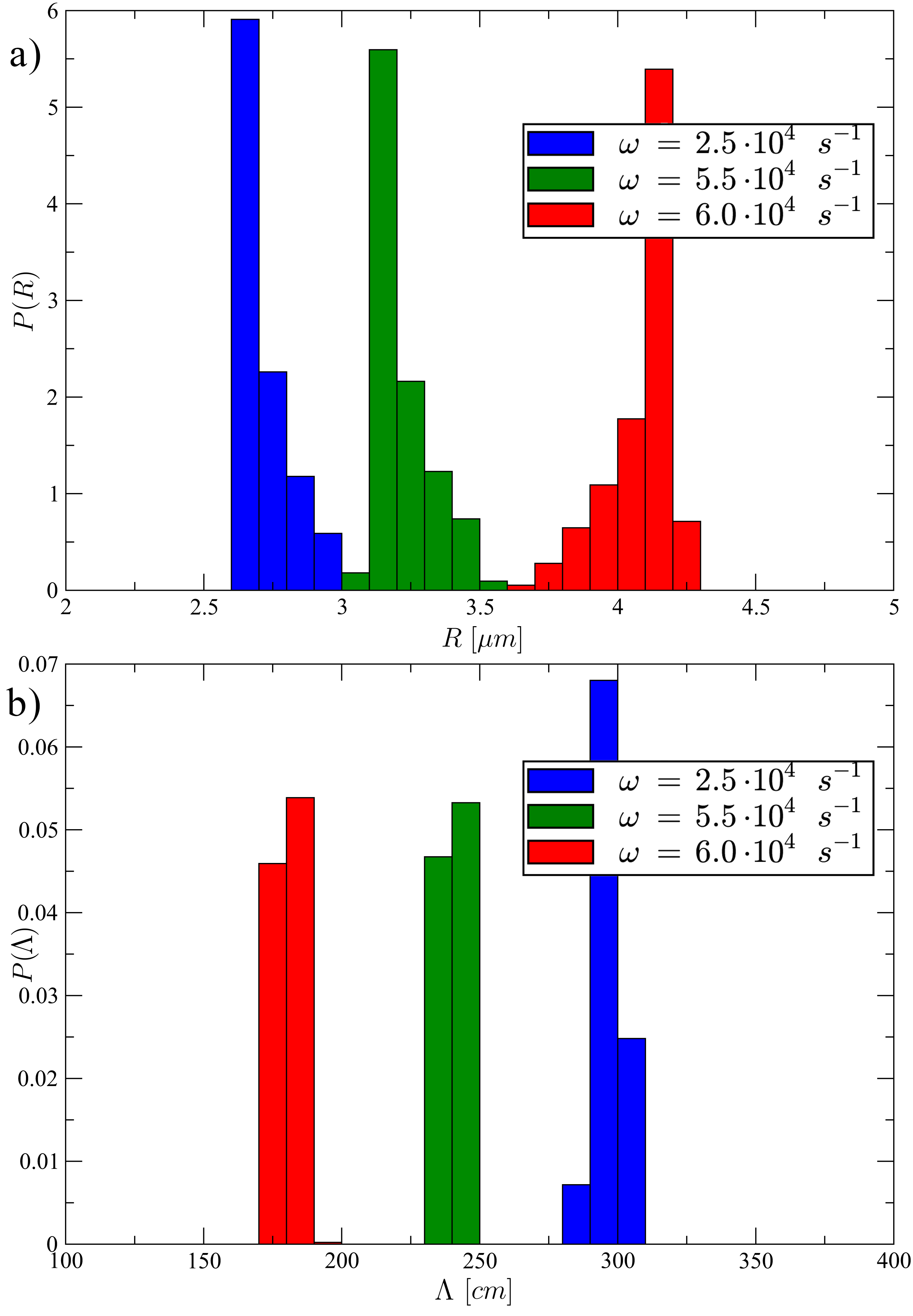}
\caption{Top: the normalized histograms for distribution of cross section radius $R$
at the collector for $A = 5$ and three values of $\omega$. 
Note that the distribution of the observable 
during the dynamics is strongly skewed and non-gaussian.
Bottom: the corresponding normalized histograms for
distribution of jet path length $\Lambda$, showing an opposite trend 
with respect to $<R>$.}
\label{fig05}
\end{figure}

The top part of Figure~\ref{fig05} shows normalized histograms for the distribution of 
cross section radius at the collector 
for three different frequencies, $\omega =[2.5, 5.5, 6] \cdot {10}^4 \, \text{s}^{-1}$, 
all with the same amplitude $A=5 \, \text{g}^{\frac{1}{2}} \, \text{cm}^{-\frac{1}{2}} \text{s}^{-1}$.
 
It is apparent that the frequency distribution is skewed and 
strongly non Gaussian, which is relatively
unsurprising due to the highly non-linear nature of the process. 
Owing to this non-Gaussianity, this
observable is best described via its median and confidence interval 
(given by the first and third quartiles),
as reported in Figure~\ref{fig03}.
In order to clarify the trend of the mean radius 
reported in Figure ~\ref{fig05}
we define the jet length as:
\begin{equation}
\Lambda\left(t\right)=\sum_{i=1}^{n-1}\left|\vec{\textbf{r}}_{i+1}-\vec{\textbf{r}}_{i}\right| ,
\end{equation}
which measures the total length of the jet, from the collector up to the nozzle. 
Comparing the two trends of $R$ and $\Lambda$ in Figure ~\ref{fig05}, it is
apparent that the two are anti-correlated, namely the smallest $<R>$ corresponds 
to the longest $<\Lambda>$ and vice versa.
As a result, one of the main effects of OREF is to alter the jet path, which in turn modifies
the jet stretch and the resulting cross sectional radius.
 
To gain further insights into the jet dynamics, it is also of 
interest to assess the "morphological distance" 
between two spirals corresponding to minimum and maximum fiber radii. 
Instead of a "smooth" Euclidean distance, we find it more informative to introduce
an overlap distance between two trajectories $\alpha$ and $\beta$ by gaining insight from an order 
parameter usually exploited in the context of glassy materials
(see Refs \cite{ozawa2015equilibrium,kob2013probing,kirkpatrick1989scaling}), which is
defined as follows :

\begin{equation}
\label{eq07}
Q_{\alpha \beta} \left( t,\varepsilon \right) =  \int_0^1 \Theta \left( \varepsilon - \vert \vec{r}_{\alpha}(t,\lambda) - \vec{r}_{\beta}(t,\lambda) \vert \right) d\lambda,
\end{equation}

In the above, $\lambda$ is the curvilinear coordinate, $\Theta$ the 
Heaviside step function, $\varepsilon$ (cm) a {\em distance} tolerance. 
The Heaviside step function acts like a switch, turning
off whenever the distance of two homologue points with the same
$\lambda$ at two different jets, goes above a given threshold $\varepsilon$. 
Therefore, $Q_{\alpha \beta}$ serves as a suitable indicator of the
{\it separation transition} between two indistinguishable ($Q=1$) 
and two fully separated ($Q=0$) configurations, at the given scale $\varepsilon$. 
Note that by assuming the jet discretized as a
set of $n$ parcels, each parcel can 
only contribute a factor $1/n$ to the overlap parameter. 
This is in contrast with the Euclidean 
distance which may eventually be completely dominated by a 
single, localized, large deviation between the two jets. 
Being sensitive to the value of $\varepsilon$,
$Q$ is a useful indicator of the dynamics 
of the separation process at different scales.

\begin{figure}[!htb]
\centering
\includegraphics[width=0.8\hsize,clip]{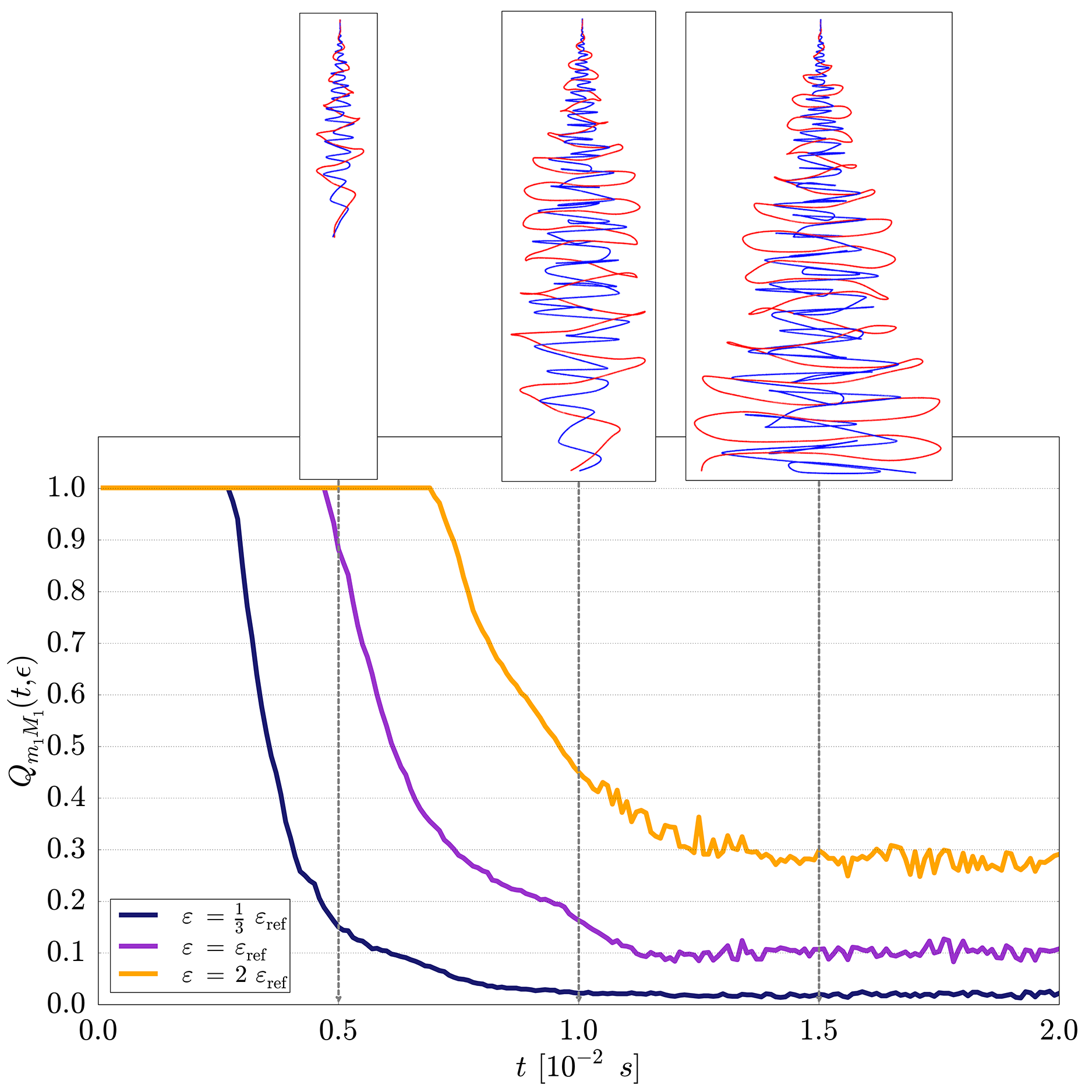}
\caption{Overlapping function $Q_{m_{1} M_{1}} \left( t,\varepsilon \right)$ 
as a function of time, for three value of $\varepsilon$. For non--zero 
values of $\varepsilon$, an abrupt transition is evident, which stops 
after the drift time $t_{\text{drift}}$ has been reached. In particular, 
the smaller the value of $\varepsilon$, the smaller $Q \left( \varepsilon,t \right)$ is. 
Snapshots of the two jets analyzed at three times are reported at the insets 
over the plot, where color legend is the same as Fig.~\ref{fig04}.
We wish to remark that the trajectories should not 
self-intersect and they actually do not.}

\label{fig06}
\end{figure}

We evaluated the overlap distance for three values of the 
parameter $\varepsilon$ as a function of time $t$,
and correlate this indicator with $R(M_{1})$  and $R(m_{1})$. 
As a reference value, we take $\varepsilon_{\text{ref}} = 0.94$ cm, i.e.  
the pitch distance of the jet-path 
in the case $m_{1}$ (see sub--panel of Figure~\ref{fig03}). 
Our results pertain to $\alpha \equiv m_{1}$ and $\beta \equiv M_{1}$, 
in Figure~\ref{fig06}, where a 
separation transition between the two jets is apparent. 
By definition, for $\varepsilon=0$,
the two jets are always separated, so that $Q_{m_{1} M_{1}}=0$ at all times.

On the other hand, at increasing $\varepsilon$, 
the two jets overlap only up
to an initial time $t_{ov}$, while for $t>t_{ov} \sim 0.01$ s, 
a separation transition starts
to take place, with the two jets getting more separated 
as  $\varepsilon$ is made smaller. After
$t = 0.15$ s, the distance between the two configurations 
reaches its asymptotic value. On top of Figure~\ref{fig06},
we report some snapshots of the two jets at a given time, 
to provide a visual counterpart
of the corresponding values of $Q_{m_{1} M_{1}} \left( t,\varepsilon \right)$.

\subsection{Time series analysis of trajectories}
\label{sec:03c}

In order to gain further insight into the jet dynamics, we analyze
the frequency spectra induced by OREF on the jet dynamics.
In particular, the Fourier analysis exposes the frequencies
of the swirling  motion of the helix, which has a direct bearing 
on the jet path and the resulting fiber radius.

To this purpose, let us consider a plane perpendicular 
to the $x-axis$ placed at $x=8$ cm,
and denote it by $\vec{\upsilon}^{*}_{\perp}=\upsilon_{y}(\lambda | x=8 \, \text{cm}) \vec{u_y}+ \upsilon_{z}(\lambda | x=8 \, \text{cm}) \vec{u_z}$ the projection 
on this plane of the velocity vector measured at the point $\lambda$ where the jet crosses such plane.

We inspect the normalized velocity autocorrelation function (VACF) 
of $\vec{\upsilon}^{*}_{\perp}$ 
defined as
\begin{equation}
Z(\tau)=<\vec{\upsilon}^{*}_{\perp}(t) \cdot \vec{\upsilon}^{*}_{\perp}(t+\tau)>_t/<\vec{\upsilon}^{*}_{\perp}(t) \cdot \vec{\upsilon}^{*}_{\perp}(t)>_t ,
\end{equation}
where bracket denote time-averaging over the corresponding steady-state regime.
The quantity $Z(\tau)$ measures the self-correlation of the swirling motion within the jet path.
Then, cosine Fourier transformation (power spectra) of the VACF was computed for each 
OREF frequency, in order to expose the spectral densities of states.\cite{allen1989computer}

\begin{figure}[!htb]
\centering
\includegraphics[width=0.8\hsize,clip]{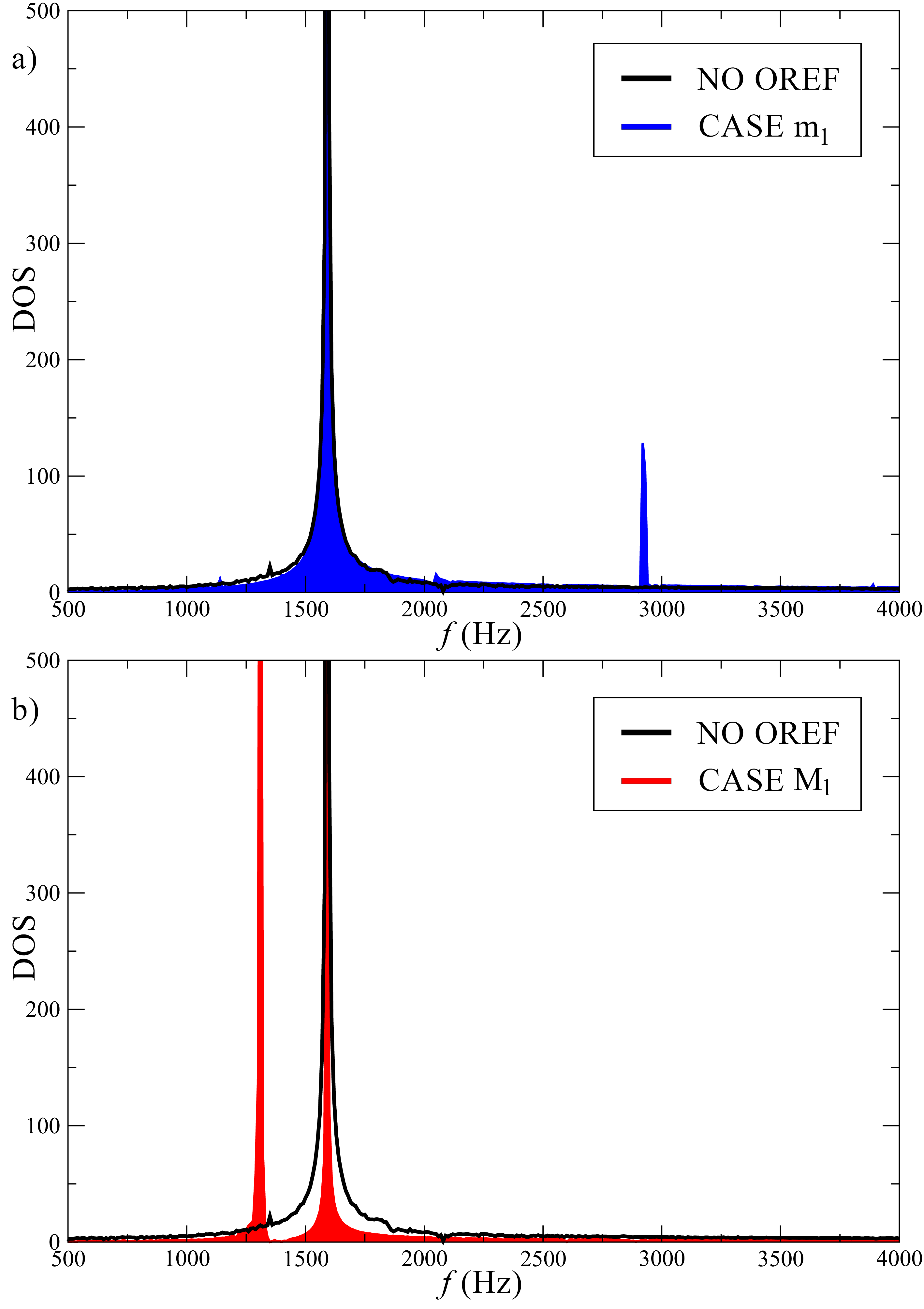}
\caption{Cosine Fourier transformation (density of states) of the VACF computed for the three cases:
without OREF as a reference case (drawn in black line), with OREF at $\omega(m_{1}) = 2.5 \cdot {10}^4 \, \text{s}^{-1}$ 
(blue curve on top panel), and $\omega(M_{1}) = 6 \cdot {10}^4 \, \text{s}^{-1}$ (red curve on bottom panel). We observe as
the case $m_1$ corresponds to the presence of a second peak located at higher frequency, while
the case $M_1$ is characterized by a second peak detected at lower frequency.}
\label{fig07}
\end{figure}

We focus our attention on three main cases: first, the simulation setup
without OREF as a reference case. The second and third correspond 
to the simulation setups denoted $m_1$ and $M_1$, respectively, in Figure \ref{fig03}.
In Figure \ref{fig07}, we observe a central peak at 1600 $Hz$ for all three cases
under inspection, which is related to the main swirling motion resulting
from the mechanical oscillation of the nozzle. 
Upon switching-on the OREF, satellite peaks are seen to appear, at higher(lower) frequencies
for the case $m_1$ and $M_1$, respectively.

In order to elucidate the relation between the jet path $<\Lambda>$ and
the frequency of the secondary peak in the power spectra, we invoke simple arguments
related to the helicoidal motion of jet.

For the sake of simplicity,  we assume (neglecting other terms as Coulomb repulsive forces, etc.) that 
the centrifugal force of the $i-th$ jet segment is approximately 
given by $F_{C,i} \propto q_{i}A$, where
$A$ is the OREF amplitude, laid on the plane of rotation, and $q_{i}$ 
is the net charge of the jet segment.
The corresponding curvature radius is $r_{c,i} \propto (q_{i}A)/(m_i \omega_{f}^{2})$, where 
$m_{i}$ is the mass of the jet segment and $\omega_{f}$ is the angular swirling frequency reported 
in Figure \ref{fig07}. 
Further, we can assess the pitch of the jet helix, defined as the height of a complete helix turn,
$h_{i}=2\pi\upsilon_{x,i}/\omega_{f}$, where $\upsilon_{x,i}$
is the velocity component of the $i-th$ jet segment along the $x$ axis.

A number of considerations on jet path and swirling frequency arise from the above 
relations. In particular, we note that the second peak in power spectra located at higher frequency
provides a reduction of the helix pitch and an increase of the jet curvature, which is
the reciprocal of the curvature radius. As a consequence, the number of helices drawn by the jet
between the nozzle and the collector significantly increases, thereby providing a larger value  
of the jet path length $\Lambda$ observed in Figure \ref{fig05}. 
Furthermore, we observe that the OREF can also be used as a tool to control 
the jet curvature, and possibly drive the jet deposition on the collector, as 
detailed in the next Subsection.

\subsection{Statistical analysis of jet deposition at the collector}

\begin{figure}[!htb]
\centering
\includegraphics[width=0.7\hsize,clip]{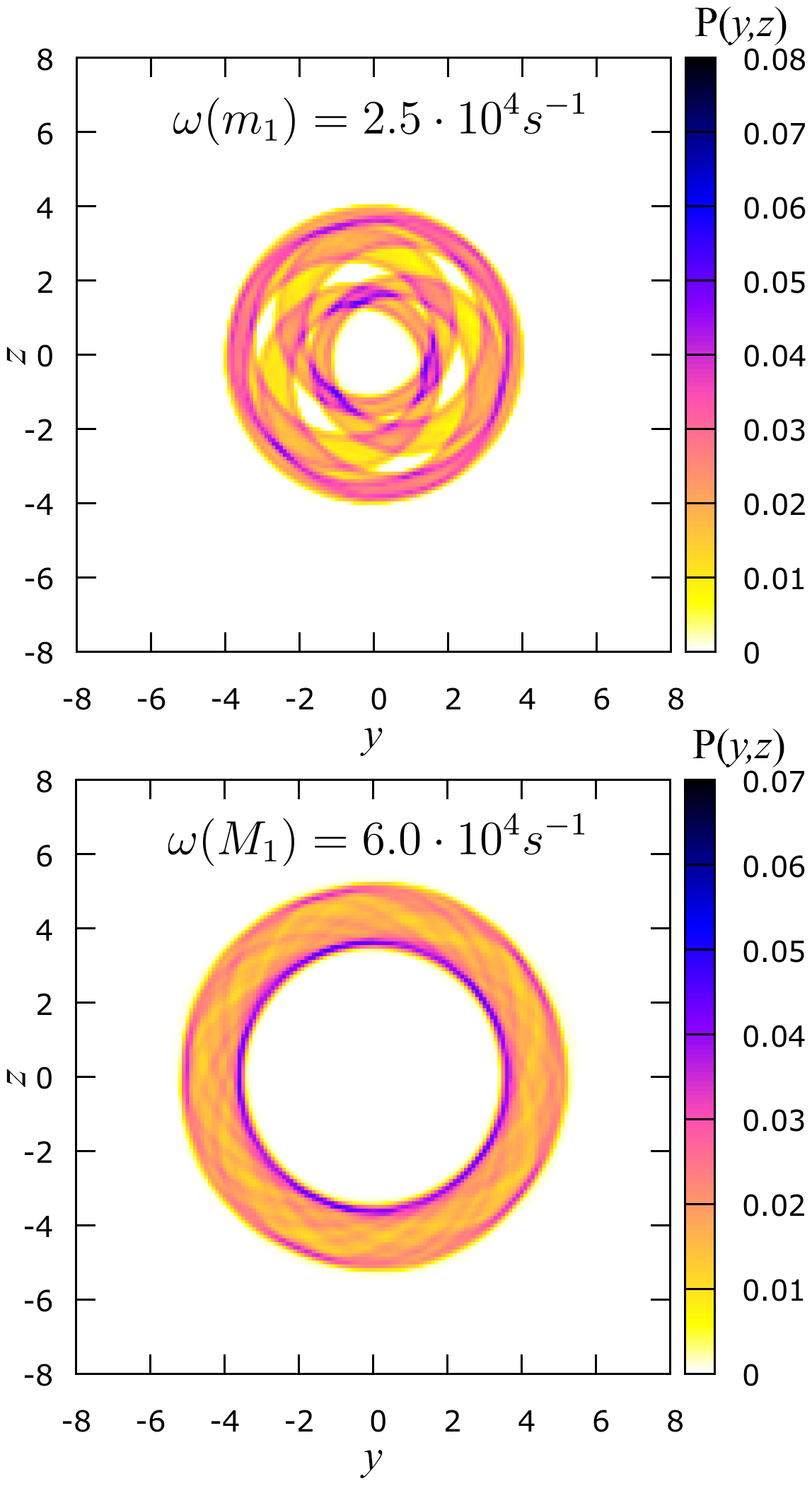}
\caption{Normalized 2D maps computed over the coordinates $y$ and $z$ of the collector for 
the two cases with OREF at $\omega(m_{1})$ (on top panel) and  $\omega(M_{1})$ (on bottom panel).
The color palettes define the probability that a jet parcel hits the collector in coordinates $y$ and $z$.}
\label{fig08}
\end{figure}

The OREF significantly affects the spatial distribution
of the jet, providing several patterns of
the electrospun coatings deposited on the collector. 
In Figure \ref{fig08}, we report the normalized 2D maps,
showing the probability of a jet parcel to hit the collector at the coordinates $y$ and $z$ 
(both perpendicular to $x$ by construction). Note that only the late dynamics 
describing the stationary regime was considered in order to compute the histograms.
Here, we observe a clear modification of the pattern deposition as
function of the applied OREF frequency $\omega$. 
In particular, by tuning the frequency, the deposition pattern is driven
towards the inner region of the collector.

This is evident in top panel of Figure \ref{fig08}, where
we report the normalized histogram for the case $m_1$, with the
probability distribution spans over a precession motion. This is likely due to the combined effects of the two peaks in
the frequency spectra (see Figure \ref{fig07}).
Since the second peak is located at higher frequency, we observe
a tighter deposition due to the larger jet curvature, depending on the frequency 
as shown in previous Subsection.
On the other hand, if the second frequency lies below the one of the
precession motion, the distribution stretches
towards the external region of collector.

Therefore, the OREF frequency $\omega$ appears to offer new, possibly
even time-dependent, strategies to achieve uniform distributions at the collector plate.

\section{Summary and outlook}\label{sec:04}

Summarizing, we have proposed the OREF mechanism and explored 
its effects on the electrospinning process, particularly on the radius of the electrospun fibers. 
Numerical simulations using the JETSPIN code show that such radius
can be reduced up to about $30 \%$. 
Despite the inherent complexity
of the underlying dynamics, the electrospinning response to 
OREF, $R = R(\omega)$ appears to organize
into a rather regular oscillatory pattern, with periodic local minima and
maxima of the finer radius as a function of the OREF frequency. 
The existence of such minima opens up the
possibility of advancing electrospinning technologies and producing 
finer fibers with high repeatability. 
Further, the OREF can be used as a control mechanism
to achieve uniform distributions of polymer filaments at the collector. 

While a $30 \%$ reduction of the diameter of single nanofibers  
might not seem that dramatic, controlling the morphology of electrospun materials 
in such a finer way, might prove useful. 
For instance, the fraction of power of the fundamental mode of a cylindrical 
waveguide strongly (exponential-like) depends on
the diameter of the guide, with a $30 \%$ variation of the diameter 
potentially leading to a significant (e.g., $20 \%$) change
of coupled optical power.\cite{fasano2013bright}
Also, fibers with such reduced size being most sensitive to their external environment, might 
lead to changes of their refractive index due
to detected chemical or biological species, which can be probed by variations in the power 
transmitted in single nanofiber waveguides. 
Finally, more uniform area coverage can lead to better coatings and improved filtrating materials.
These perspectives are intriguing and worth experimental investigation.

The present simulations permit to highlight the salient morpho-dynamical 
features associated with the application of the OREF, as well as to probe
the electrospinning response in a range of applied frequency.
Much remains to be done for the future; particularly, 
the study of the spatial dependence of the 
self-consistent electrostatic field induced by charge 
deposition at the collector, and its effects 
on the overall jet dynamics and associated deposition patterns. 
Moreover, a fully-fledged analysis of the non-linear dynamical behavior
of the OREF setup would be highly desirable.
Studies along these lines are currently under way.

\begin{acknowledgments}
The research leading to these results has received funding from the
European Research Council under the European Union's Seventh Framework
Programme (FP/2007-2013)/ERC Grant Agreement n. 306357 (NANO-JETS).
\end{acknowledgments}

\end{document}